\documentclass{article}
\usepackage{graphicx} 
\usepackage[utf8]{inputenc}
\usepackage{float}
\usepackage{amsmath}
\usepackage{setspace}
\usepackage{upgreek}
\usepackage{siunitx} 
\usepackage{graphicx}
\usepackage{amsmath, amssymb}
\usepackage{authblk}
\usepackage{url}
\usepackage{caption}
\usepackage[margin=4cm]{ geometry }

\title{Dynamic Scattering-channel-based Approach for Multiuser Image Encryption}

\date{ }
\author[1] { Mohammadrasoul Taghavi}
\author[1] { Edwin A. Marengo}
\affil[1]  { Department of Electrical and Computer Engineering, Northeastern University, Boston, Massachusetts 02115, USA}

\begin{document}
\maketitle

\noindent
{\bf {\Large Abstract}}:
Conventional scattering-based encryption systems that operate based on a static complex medium which is used by all users are vulnerable to learning-based attacks that exploit ciphertext-plaintext pairs to model and reverse-engineer the scattering medium's response, enabling unauthorized decryption without the physical medium. In this contribution, a new dynamic scattering-channel-based technique for multiuser image encryption is developed. The established approach employs variable, dynamic scattering media which are modeled as tunable aggregates of multiple scattering nanoparticles. The proposed system supports multiple users by allowing distinct combinations of scattering  matrices for different time blocks, each combined with user-specific complex-valued coefficients, enabling the creation of unique, hard-to-guess encryption keys for each user. The derived methodology enhances the practical feasibility of multiuser secure communication and storage channels employing scattering media as the encryption mechanism. 
\section{Introduction}

The encryption of images in the realm of secure communication and storage of information has emerged as an exciting area of research \cite{liu2014review, kumari2017survey, chen2010optical,zhang2008optical, li2015compressive, taghavi2024differential}. Recently, the use of complex scattering media for image encryption has attracted considerable attention due to its ability to provide high levels of security through scrambling of information \cite{liu2020exploiting,bai2022coherent, zhao2022speckle, yu2023ultrahigh}. These methods rely on transforming an image into a speckle-like pattern or hologram, which acts as the ciphertext. The correct decryption of this ciphertext, for revealing the original image or plaintext, requires the physical presence of the appropriate scattering medium or an accurate model of it. The scattering medium, or its model, serves as the encryption key, with the medium's physical properties playing a crucial role in the overall security mechanism. Researchers have explored various approaches by employing both static and dynamic scattering media, and leveraging their inherent randomness for imaging purposes. For example, Ruan et al. \cite{ruan2020fluorescence} introduced a technique that combines ultrasound and light to achieve high-resolution fluorescence imaging through dynamic scattering media. By utilizing time-varying speckle patterns, this approach addresses the challenges of fast speckle decorrelation, enabling improved imaging resolution compared to conventional optical methods. In another important work, by utilizing time-reversed ultrasonically encoded light, Liu et al. \cite{liu2015optical} developed a method that enables optical focusing in dynamic scattering media. This approach tackles the difficulty of concentrating light within highly scattering environments, opening up possibilities for noninvasive deep-tissue imaging applications. Wang et al. \cite{wang2021non} introduced a non-invasive super-resolution imaging technique that captures speckle patterns from blinking point sources, enabling high-resolution imaging through dynamic scattering media. 

An important application of scattering-based encryption lies in public encryption systems where the same physical encryption apparatus or server is shared by a multitude of users, each seeking encryption of their individual signals, for storage or communication purposes. Within this important practical framework, a number of specific challenges arise. 
For instance, if the system is based on static complex media, then the shared nature of the channel introduces potential vulnerabilities. Despite the promising security offered by static scattering media in encryption, recent advancements in deep learning have exposed potential vulnerabilities \cite{wang2021holographic, yan2024image}. Using exhaustive queries by intruders, machine learning algorithms can be trained on pairs of plaintext and ciphertext to model the scattering medium's response fully (or partially), effectively reverse-engineering the encryption process \cite{zhou2020learning,zhang2023speckle}. 
This ability to approximate the scattering medium's behavior through repeated observations can undermine the security of systems that rely on a static medium. As a result, adversaries (e.g., masquerading as valid users) may be able to extract (through exhaustive queries) the encryption key, thereby putting the legitimate users' sensitive information at risk. 

In this paper, we propose a new multiuser encryption technique based on sets of dynamic scattering matrices and their combinations. 
The proposed method involves the adoption, within each time block, of plaintext-to-ciphertext transformations that are based   on a large set of variable, dynamic scattering states.
The scattering states are constantly shuffled from a block to another, and a
unique combination of these states, as governed by a randomly generated (and thereby probabilistically unique) complex-valued weight or expansion coefficient, is allocated to each user. 
This scheme, comprising 1) scattering state multiplicity per block, 2) random shuffling per block, as well as 3) unique user-dependent linear combinations per block significantly increases the corresponding key space, thereby making it 
quite challenging for attackers to break the encryption through the aforementioned brute force and reverse-engineering methods. 
Moreover, modulation techniques such as phase shifts or scaling can also be incorporated to further differentiate the encryption keys used by different users. In this research, we model the envisioned scattering media as tunable conglomerates of  multiple scattering nanoparticles whose physical realization aligns, e.g., with recent advances in nanoparticle-related technologies \cite{shirmanesh2020electro,cui2019tunable,taghavi2024active}. Time-varying meta-atoms, for example, are highly relevant. They enable dynamical control of the scattering response of nanoantennas \cite{sisler2024electrically,barati2020time, barati2022optical, sadafi2023dynamic}. These innovations allow for real-time manipulation of light in metamaterial-based interfaces and volumes by engineering the scattered light's behavior \cite{barati2022optical2, mohammadi2023active}. 

The rest of the paper is organized as follows. The proposed encryption method, which is based on an inverse scattering scheme, is explained in detail in section 2 within the context of the envisaged multiuser encryption and communication applications. Section 3 presents the results of computer simulations illustrating the feasibility of the proposed encryption approach. Section 4 provides concluding remarks. 
 \begin{figure}[H]
    \centering
    \includegraphics[width=12 cm]{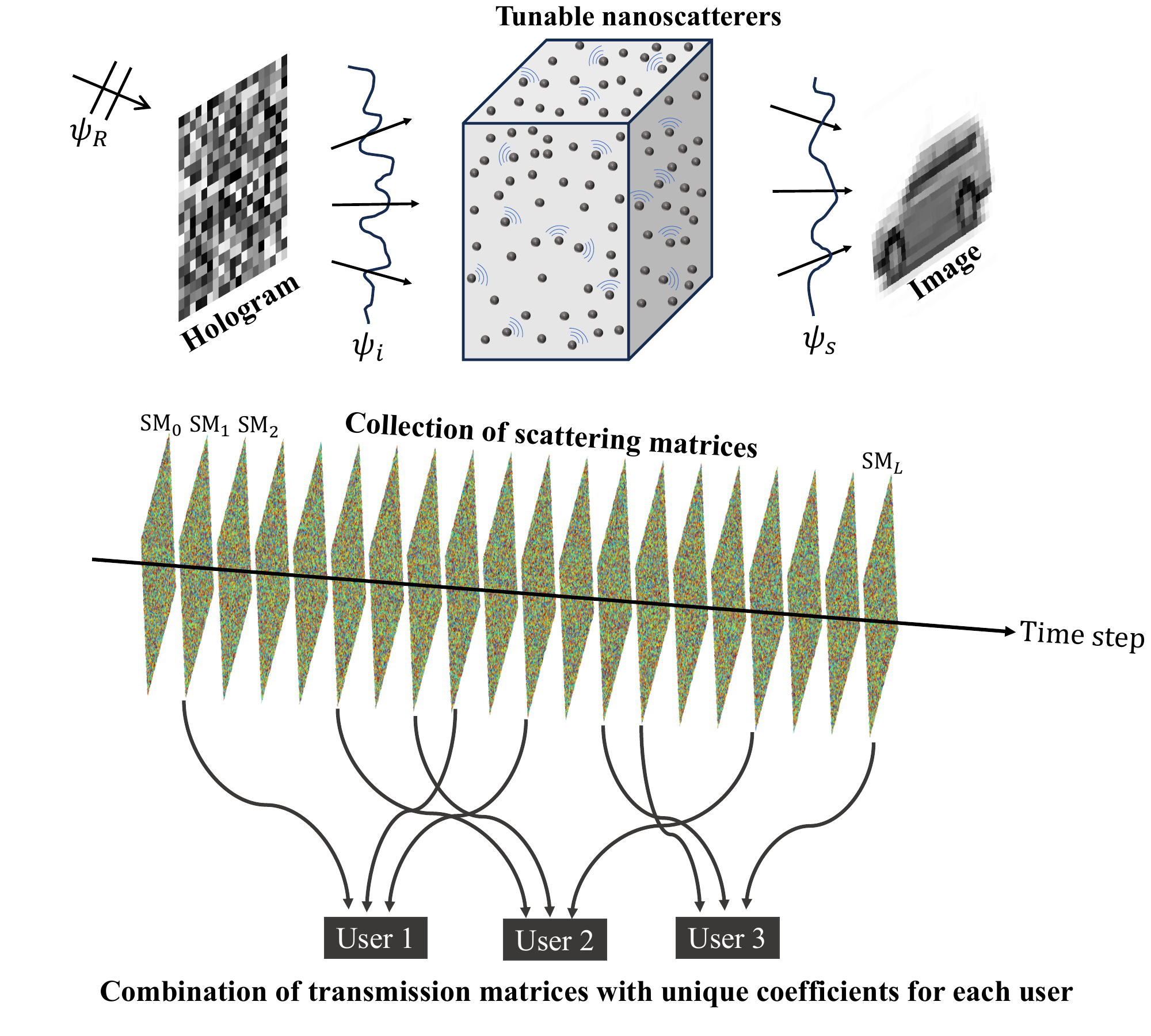}
    \caption{Schematic of the proposed dynamic scattering-based encryption system.}
    \label{fig_1}
\end{figure} 

\section{Proposed Encryption System}
Scattering-based encryption methods can be based on either forward or inverse scattering mechanisms. In the forward method, the information-carrying object whose image one seeks to encrypt is adopted as input or illumination, and the resulting scattered signal is used at the receiving end for the generation of the ciphertext, ideally in the form of a hologram to store the pertinent phase information. In this framework, incident field $\psi_i$ is produced upon illumination of the object and scattered field 
\begin{equation}
    \psi_s = K \psi_i \label{eq0_sept14_2024}
    \end{equation}
    is generated through the interaction of the probing field with the complex scattering medium whose response is characterized by scattering matrix $K$. Then intensity-only ($I$) or holographic ($H$) measurements are stored, giving rise to the ciphertext, in particular, 
    \begin{equation}
        I(n) = |\psi_s({\bf R}_n)|^2  \quad n=1,2,\cdots, N \label{eq1_sept14_2024}
    \end{equation}
    or the hologram 
    \begin{equation}
        H(n) = |\psi_s({\bf R}_n) + \psi_R ({\bf R}_n)|^2 \quad n=1,2,\cdots, N \label{eq2_sept14_2024}
    \end{equation}
where $\psi_R$ denotes a reference wave and ${\bf R}_n, n=1,2,\cdots, N$ represent $N$ sampling points or pixels within a suitable sensing plane $\Sigma$. In the decryption phase, an inverse problem is solved, either computationally (e.g., via machine learning \cite{zhao2022speckle}) or in hardware via complex conjugation \cite{popoff2010image}, and an image that approximates the sought-after object is obtained. 
In this method the object or plaintext information is carried by the incident wave $\psi_i$ while the ciphertext is based on the scattered field $\psi_s$. 

In the complementary approach the roles of $\psi_i$ and $\psi_s$ are reversed, and thus the ciphertext (e.g., a suitable hologram) carries the information needed for the launching of an incident field $\psi_i$ whose corresponding scattered field $\psi_s$ is an image of the object of interest (see Fig.~1). Thus in this alternative methodology a ciphertext hologram is designed, for the relevant encryption scattering medium, via inversion of  the forward map in (\ref{eq0_sept14_2024}), such that upon illumination with a suitable reference wave ($\psi_R$) a desired incident field $\psi_i$ is launched, whose corresponding scattered field $\psi_s$ is approximately equal (within applicable resolution limits) to the desired information-carrying image. This inversion can be done computationally, based on real or synthetic data. In the latter case the resulting ciphertext hologram is purely computationally-generated (CGH).

To elaborate on the required inversion, we assume that the hologram in question has $M$ pixels, while (as in (\ref{eq1_sept14_2024}) and (\ref{eq2_sept14_2024})) the scattered field is captured using $N$ pixels. We consider in the following the regularized pseudoinverse of the forward map $K$ which we express in terms of the (truncated) singular system of $K$ as 
\begin{equation}
    \hat\psi_i = \sum_{p=1}^{P_0(\varepsilon)} \sigma_p^{-1} \psi_i^{(p)} \left( \psi_s^{(p)} \right)^{\dagger} \psi_s \label{eq3_sept14_2024}
\end{equation}
where  $\dagger$ denotes complex conjugation while $\sigma_p, p=1,2,\cdots, P$ are the singular values, where $P=\min(M,N)$ while $\psi_i^{(p)}$ and $\psi_s^{(p)}$ are the right-hand (input) $M\times 1$ singular vectors and left-hand (output) $N\times 1$ singular vectors, respectively, while $P_0(\varepsilon)\leq P$ is the number of significant singular values, such that $\sigma_p>\varepsilon$ where $\varepsilon$ is a regularization parameter that is chosen in a tradeoff between inversion stability and accuracy, as is well known (\cite{bertero2021introduction}, ch.~5) Then the desired output image is represented by $\psi_s$ and the required incident field $\hat\psi_i$ is computed via (\ref{eq3_sept14_2024}). The next step is the determination of the corresponding ciphertext hologram. The 
 required hologram transparency $t$ is given by
 \begin{equation}
      t ({\bf X}_m) = | \hat\psi_i({\bf X}_m)+\psi_{R}({\bf X}_m)|^2 \quad m=1,2,\cdots, M  \label{eq_aug_8_2024_15}
 \end{equation}
 where ${\bf X}_m, m=1,2,\cdots,M$ denote the pixel positions in the ciphertext hologram plane. 
When illuminated by the reference wave $\psi_{R}$, this transparency produces four field components, one of which corresponds to the virtual image field, matching the desired incident field $\hat\psi_i$. By employing an appropriate design, such as Leith-Upatnieks holography, the influence of the other field components can be minimized near the scattering medium, thereby enabling (in the decryption phase) the approximate reconstruction, at the sensing plane $\Sigma$, of the sought-after waveform $\psi_s$ (plaintext).

These results apply for each realization of the scattering matrix $K$. We discuss next the procedure through which different scattering matrices are randomly constructed and allocated to different users. We consider the general scenario in which there are multiple transmitters and receivers in the channel, labelled $q=1,2,\cdots, Q$, where, e.g., the $q$th user wishes to securely transmit an image to the $q'$th user through the scattering-based encryption which we envision as a form of cloud encryption. In this context, user $q$ uploads the plaintext image to the encryptor which generates a unique random key (complex coefficients $C(q,q')$) associated to the process (communication from $q$ to $q'$) along with the corresponding ciphertext. Then the encryptor sends the ciphertext back to the transmitter, which can send it to the intended recipient via a communication link. A unique identification receipt or key is also sent to the intended receiver (user $q'$). Upon reception, the intended recipient can upload the ciphertext into the encryptor, alongside the pertinent identifying credentials to validate this user's identity as the intended $q'$th user. The encryptor then retrieves from the ciphertext, via the unique process-linked key $C(q,q')$, the sought-after plaintext which is finally released to user $q'$, as intended. 
Figure 1 shows a schematic of the envisaged system, within the inverse scattering framework. The scattering medium varies dynamically, taking a sequence of $L$ scattering states corresponding to scattering matrices $SM_1, SM_2, \cdots, SM_L$ within each block or period of the communication link. These states are shuffled randomly from a block to another, in a sequence that is kept secret from the users (hidden variable). Given $Q $ users in the channel, a unique combination of said matrices is assigned per communication block for all the active user pairs. In particular, the scattering response $K_{q,q'}$ assigned to user pair $(q,q')$ is of the form
\begin{equation}
    K_{q,q'}= \sum_{l=1}^L C(q,q';l) (SM_l) \label{eq_7_sept14_2024} \quad q=1,2,\cdots,Q; q'=1,2,\cdots,Q
\end{equation}
where the set of complex-valued expansion coefficients $C(q,q';l)$ is chosen non-adaptively, randomly, for each user pair. The case $q=q'$ is an important special case in which a user seeks to encrypt information for secure storage, e.g., in situations where it must be stored in a potentially hackable medium or server. In this application, only said user can subsequently recover the plaintext, via access to the aforementioned transaction receipt. Another special version of (\ref{eq_7_sept14_2024}) is shown in Fig.~1, wherein the sought-after unique combination is achieved via the selection of a unique subset of the scattering matrices for each user, alongside a unique superposition of the form (\ref{eq_7_sept14_2024}) for the matrices in each subset. 

In the envisioned application, user $q$ uploads within each block an information-carrying image (say $\psi_{s}^{(q)}$) whose encryption is sought. The intended receiver ($q'$) is indicated (via a password, or another credential). The encryption system randomly assigns the corresponding process-linked key $C(q,q';l)$ and carries out the inversion in (\ref{eq3_sept14_2024}) for $K\rightarrow K_{q,q'}$, so as to generate the required user-dependent probing field (say $\hat\psi_i^{(q\rightarrow q')}$) and associated ciphertext hologram (say $t^{(q\rightarrow q')}$; see (\ref{eq_aug_8_2024_15})). The encryption server stores the keys ($C(q,q';l)$ and scattering states ($SM_l,l=1,2,\cdots, L$)) required for subsequent retrieval of the plaintext image. At this point, user $q$ sends ciphertext hologram $t^{(q)}$ to user $q'$. Finally, upon a validated request by user $q'$, plus communication of ciphertext hologram $t^{(q\rightarrow q')}$, the encryption server generates, for each relevant state ($SM_l, l=1,2,\cdots,L$) the corresponding scattered field which is stored in the form of a hologram similar to $H$ in (\ref{eq2_sept14_2024}). We shall denote these holograms as $H^{(l)}, l=1,2,\cdots,L$. The retrieval of an information-carrying plaintext hologram, to be denoted as $h^{(q\rightarrow q')}$ to emphasize communication from $q$ to $q'$, is achieved from (\ref{eq_7_sept14_2024}) via
\begin{equation}
    h^{(q\rightarrow q')}=\sum_{l=1}^{L} C(q,q';l) H^{(l)} .\label{eq1_sept16_2024}
\end{equation}
This hologram is communicated to the intended receiver $q'$ for the synthesis of the sought-after image of $\psi_s^{(q)}$
through illumination of the hologram with the pertinent reference wave $\psi_R$, 
thereby completing the cycle of encryption, secure communication, and decryption, as desired. In the next section we present  computer validations for the special case where $q=q'$, corresponding to scenarios where the cloud encryption tool is used for individual secure data storage. 
\section{Computer Simulations}
\begin{figure}[H]
    \centering
    \includegraphics[width=14 cm]{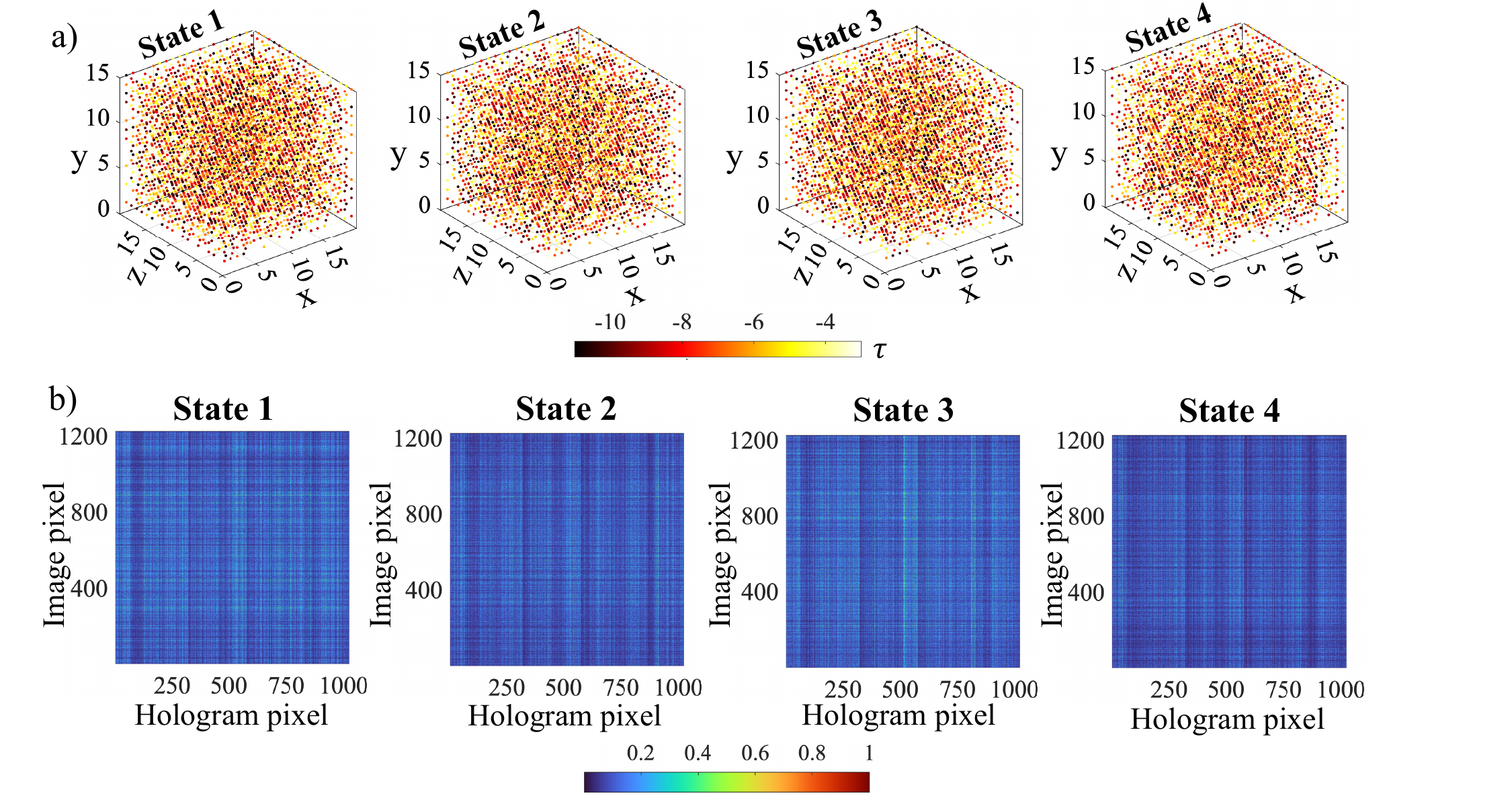}
    \caption{(a) Graphical representation of four aggregates of scattering nanoparticles, corresponding to the four distinct states adopted in the simulations, and (b) amplitude of the corresponding scattering matrices.}
    \label{fig_2}
\end{figure}

In this section, we demonstrate a numerical example for the proposed dynamic scattering-channel-based encryption based on scalar wave scattering simulations in three-dimensional (3D) space. Both the hologram and image regions are defined in two dimensions (2D), with the ciphertext hologram having dimensions of $64\lambda \times 64\lambda$ and pixel sizes of $n_x = 35$ and $n_z = 35$. The image sensor, on the other hand, measures $70\lambda \times 70\lambda$ with pixel sizes of $n_x = 32$ and $n_z = 32$. The distance between the ciphertext hologram and the image sensor is set to $70\lambda$. Also, the distances between the scattering medium and the image sensor, and between the medium and the hologram, are both $10 \lambda$. The scattering medium is a 3D volume with dimensions of $70\lambda\times 70\lambda \times 50\lambda$ along the x, z, and y axes, with the y-axis representing the propagation direction. The considered medium consists of $n_x=20$, $n_y=15$, and $n_z=20$ scattering centers or nanoparticles. 
In the simulations, the particle scattering strengths or potentials were randomly selected between -3 and -11. We considered the canonical case of four scattering states within a functional block of the encryption system (corresponding to $L=4$ in the preceding analysis). The configurations used in the simulations, which were randomly selected, are illustrated in Figure~\ref{fig_2}(a).  
The associated scattering responses or matrices were computed using the Foldy-Lax multiple scattering method \cite{ishimaru2017electromagnetic,huang2013efficient}, as full-wave techniques are computationally expensive and inefficient for this task. Plots of the normalized amplitude of the resulting
scattering matrices
are shown for reference in Fig.~\ref{fig_2}(b). The responses for these configurations were found to be quite different, as desired, in order to minimize cross-talk. 

To study the performance of the proposed encryption method, we utilized images from the CIFAR-10 dataset, which consists of 60,000 32x32 color images across 10 different classes and is widely used for image classification and machine learning research \cite{krizhevsky2009learning}.
We converted the selected colored images to grayscale in order to simplify the simulations. We considered the case of three users ($Q=3$) who seek to securely encrypt their information (in the present case, three (plaintext) scattered field images $\psi_s^{(1)}$, $\psi_s^{(2)}$, and $\psi_s^{(3)}$ (see Fig.~3)) through the aforementioned four-scattering-state encryption system.  
Figure 3 shows the results of a representative simulation. The first row shows the plaintext images of the three users (corresponding to $\psi_s^{(1)}$, $\psi_s^{(2)}$, and $\psi_s^{(3)}$, for users $1$, 2, and 3, respectively). Following the discussion in (\ref{eq_7_sept14_2024}), unique scattering matrices were allocated to the users, which we denote as $K_1$, $K_2$, and $K_3$, for users $1,2$ and $3$, respectively. In this example we examined encryption via the following unique combinations of the available scattering matrices: $K_1 = a(1) TM_1 + b(1) TM_2 + c(1) TM_3$ , $K_2 = a(2)TM_1 + b(2) TM_2 + c(3) TM_4 $ ,  and $K_3 = a(3) TM_2 + b(3) TM_3 + c(3) TM_4$ where $a(q),b(q),c(q), q=1,2,3$ are unique user coefficients of the form $(a,b,c)=(\exp{(-jC_a\pi)},\exp{(-jC_b\pi)},\exp{(-jC_c\pi)})$ (where in the latter equation, and in the following discussion, the $q$-dependence of these coefficients is suppressed for clarity). In the simulation under consideration $(C_a,C_b,C_c)$ were set to (-0.3,0.7,-0.8), (-0.6,0.2,-0.8), and (-0.4,0.7,0.3) for users 1,2, and 3, respectively. As explained in the previous section, a unique ciphertext hologram is developed for each user, according to the given plaintext and the corresponding unique scattering matrix ($K_q, q=1,2,3$). Upon excitation with the relevant reference wave, the hologram gives rise to an incident field whose corresponding scattered field for transmission matrix $TM_l$ is denoted as $\psi_l, l=1,2,3,4$. The figure shows images of these scattered fields for each user which, as expected, have speckle-like characteristics and bear no resemblance to the sought-after plaintext image. Moreover, the image corresponding to the sum of the individual scattering matrices' images also lacks similarity to the intended user image. The figure illustrates that only when the correct user-allocated keys are selected, corresponding to a unique combination of the same fields ($\psi_l, l=1,2,3,4$), does the image reveal the correct plaintext uploaded by the user. 
\begin{figure}[H]
    \centering
    \includegraphics[width=12cm]{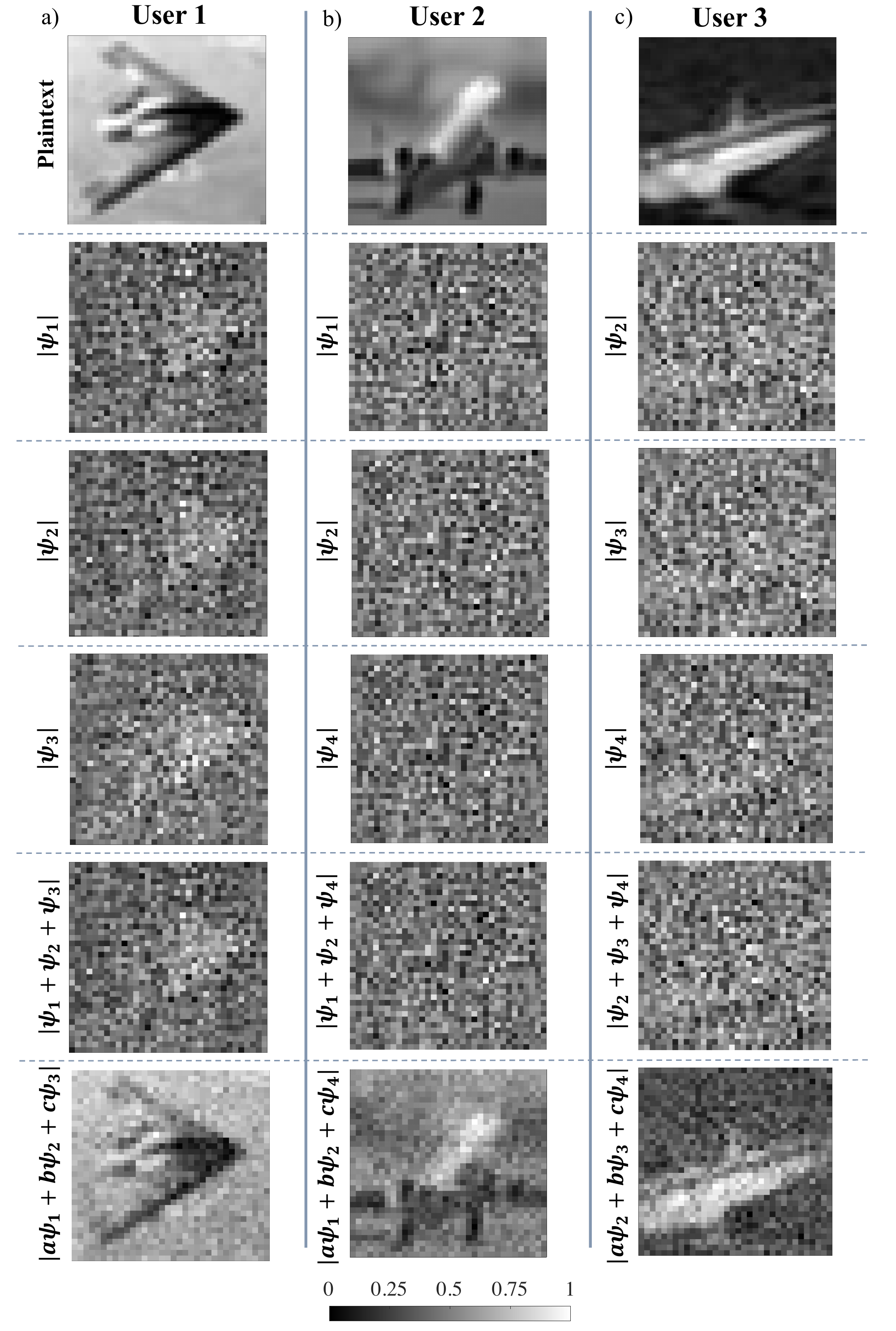}    
\caption{Illustration of the dynamic scattering-based encryption scheme for three different users: (a), (b), and (c). The figure shows the images corresponding to reconstruction attempts based on different combinations of the scattered signals $\psi_1$, $\psi_2$, $\psi_3$, and $\psi_4$ corresponding to block scattering matrices $TM_1$, $TM_2$, $TM_3$, and $TM_4$, respectively. Successful reconstruction is achieved only when the correct user-assigned keys are used.}
\label{fig_3}
\end{figure}
  These results are encouraging and are representative of what we observed in several other cases explored in the course of this research (results not shown). It is important to point out that the quality of the reconstructed image is not perfect due to intrinsic imaging limitations linked to the adopted scattering system setup where, in particular, both the hologram and image planes are separated from the medium by relatively large distances ($>>2\lambda$), thereby rendering loss of evanescent information, while at the same time both apertures are space-limited and the data are handled in a sampled manner thereby causing additional computational limitations. 
  \begin{figure}[H]
    \centering
    \includegraphics[width=12 cm]{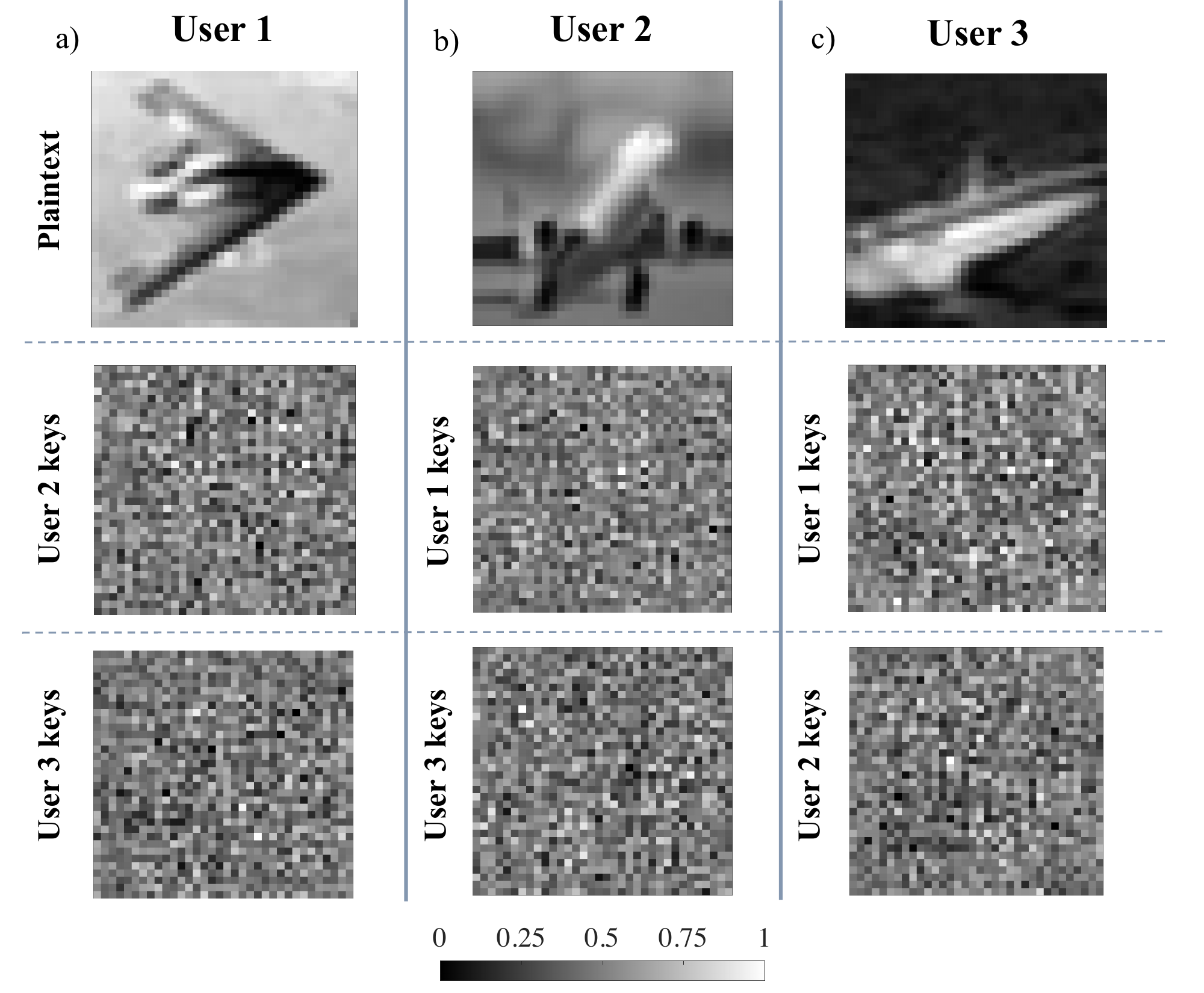}
    \caption{Analyzing the potential vulnerability of the system to cross-channel interference, where keys from other users are used to reconstruct the original images (a), (b), and (c) of each user. The analysis suggests that no meaningful information about the plaintext images can be extracted using the keys of other users.}
    \label{fig_4}
\end{figure} 
 
\begin{figure}[H]
    \centering
    \includegraphics[width=12 cm]{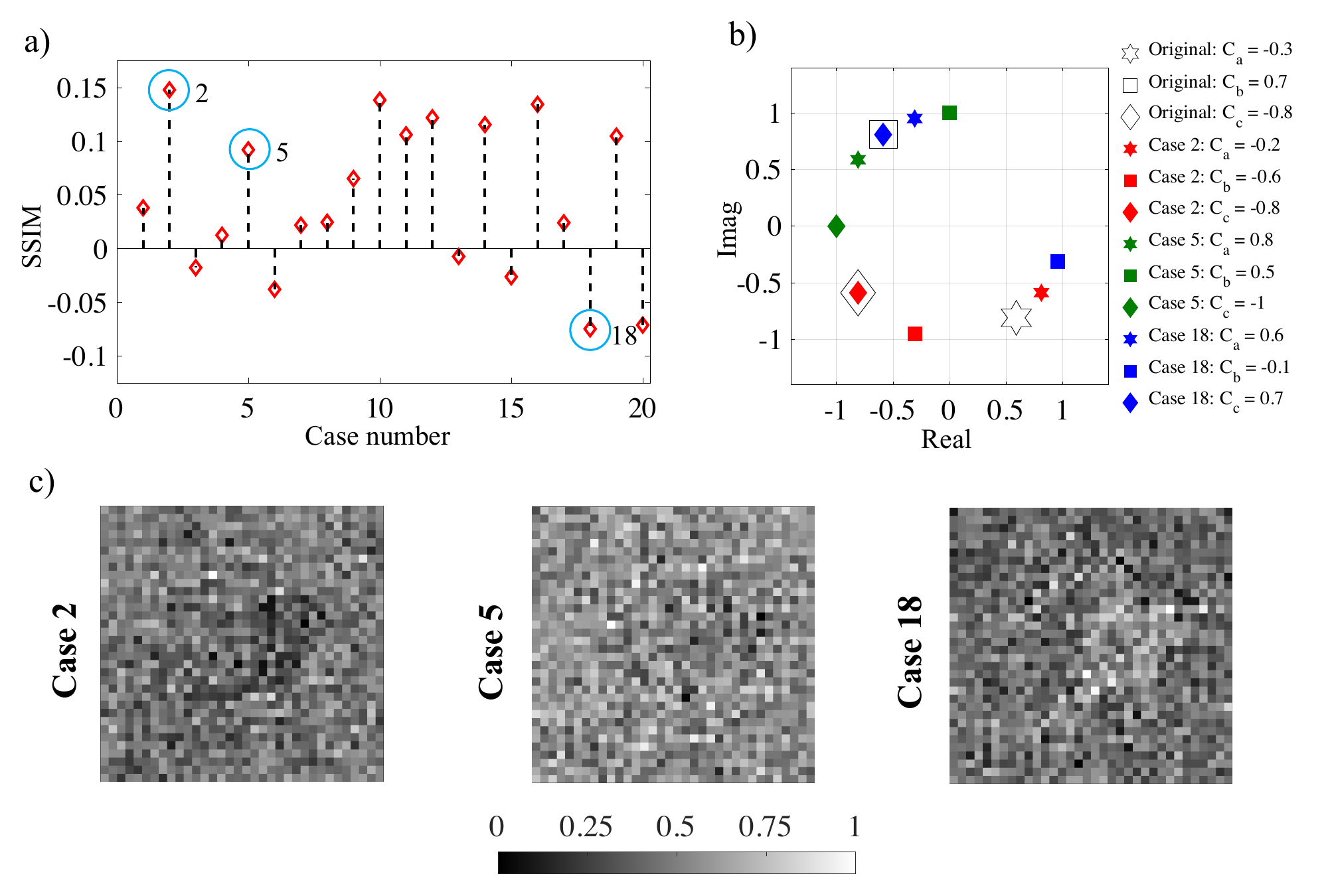}
    \caption{Security analysis of the proposed encryption system. (a) SSIM calculation for twenty random key combinations attempting to hack user 1's information, assuming the intruder has knowledge of the specific scattering matrices used in the hologram generation process. (b) Correct and selected fake keys, represented as $(a, b, c) = (\exp{(-jC_a\pi)}, \exp{(-jC_b\pi)}, \exp{(-jC_c\pi)})$, for cases 2, 5, and 18, are plotted on the complex plane. (c) Visual representation of the reconstructed image for the selected three cases to get a better understanding of the system performance.}
    \label{fig_5}
\end{figure} 

 To further evaluate the security of the proposed encryption method, we conducted an additional set of simulations to determine whether the plaintext image of a given user could be partially reconstructed using keys assigned to the other users. This simulates the situation in which the non-intended user gains access to the ciphertext of another user and attempts to reconstruct using the available key. Each column in Fig.~\ref{fig_4} presents the results of the these simulations. As shown, incorrect reconstructions are obtained with the other users' keys and clearly the encryption channel remains secure against this type of attack. 

 To enable further numerical and visual comparison of the reconstruction quality across different scenarios, we also employed the structural similarity index (SSIM), a well-known metric in image processing \cite{nilsson2020understanding}.
It offers distinct advantages over traditional metrics like mean squared error (MSE) and peak signal-to-noise ratio (PSNR) by considering perceptual aspects such as luminance, contrast, and structural information. Unlike MSE, which focuses on pixel-level differences, SSIM evaluates how well the reconstructed image maintains the overall structural and visual coherence, making it more aligned with human visual perception. This makes SSIM particularly effective for assessing image quality in tasks like encryption and decryption, where structural fidelity is more significant than minor pixel variations. Furthermore, SSIM produces a value between -1 and 1, where an SSIM of 1 indicates a perfect reconstruction, while values approaching zero reflect a decline in reconstruction quality. Negative values suggest that the reconstructed image lacks any meaningful similarity to the reference.

One specific intrusion scenario to consider is when an eavesdropper gains access to both the hologram and the scattering matrices used in the key. The eavesdropper, however, does not know the relevant user-allocated coefficients required for the reconstruction. In that case, the eavesdropper may adopt various combinations of $(a,b,c)$ to computationally reconstruct the output. Exploring this scenario is crucial for assessing the robustness of the system against such attacks. Representative results are shown in Fig.~\ref{fig_5}(a), which shows the SSIM value corresponding to twenty random combinations or keys. The obtained maximum SSIM for such randomly generated keys is around 0.15. In comparison, the SSIM calculated for the originally reconstructed signal is 0.64. Additionally, Fig.~\ref{fig_5}(b) illustrates, in the complex plane, the location of the correct key and of three pairs  of selected fake keys ($a,b,c$) (corresponding to cases 2, 5, and 18 in Fig.~\ref{fig_5}(a)). The complex plane plot in Fig.~\ref{fig_5}(b) reveals that the highest SSIM (case 2) corresponds to a scenario where the intruder intercepts one correct key component, $C_c$. Additionally, in that case the intruder has nearly obtained another key component (in particular, $C_a$) making this a particularly critical and risky situation. However, even in this extreme case, where two of the key components have been compromised, the reconstructed signal still bears minimal resemblance to the original plaintext image, demonstrating the robustness of the encryption system. This is quite encouraging and remarkable, particularly since these results hold for a very small number of scattering states ($L=4$). The encryption performance is expected to improve exponentially as the number of such dynamic states increases, of course. Finally, to provide a clearer understanding of what an SSIM value of 0.15 represents in terms of visual perception, we plot in Fig.~\ref{fig_5}(c) the reconstructed images for the aforementioned three cases (2, 5, and 18). It is evident that in none of the cases the reconstructed image provides any discernible information about the valid user's data. Similar results (not shown) were obtained for other images from the mentioned dataset, as well as for other scattering configurations and user coefficients, thereby corroborating the validity of the derived dynamic scattering-channel-based encryption technique. 
\section{Conclusions}
We have demonstrated the feasibility of utilizing dynamic scattering matrices for the secure encryption of information such as images uploaded by and communicated to multiple users employing the same encryption server. In this scheme, each user is assigned a unique combination of scattering matrices with specific coefficients. This approach makes it significantly harder for attackers to break the encryption, in comparison with more conventional architectures where all users rely on a single static scattering matrix per encryption block. The proposed dynamic evolution of scattering matrices, including shuffling effects, further strengthens the associated encryption security in the envisaged multiuser environments. The developed approach was illustrated with a canonical example involving three users and a rather simple encryption server based on only four scattering states per encryption block. In this example, meaningful reconstructions were achieved for each user only with the adoption of the correct user-assigned keys, as expected. The obtained results are encouraging and remarkable, particularly taking into account that they involve only four states. More formidable results may be achievable, of course, for larger state numbers corresponding to more realistic practical implementations of this approach. 

\section*{Acknowledgments}
The authors wish to thank Northeastern University’s Research Computing team for providing resources for high-performance computing through Discovery Cluster.
\bibliography{main}
\bibliographystyle{unsrt}
\end{document}